%% file: main.tex
\documentclass[preprint,times]{elsarticle}

\usepackage{hyperref}
\usepackage{siunitx}
\usepackage{booktabs}
\usepackage{float}
\usepackage{color, soul}
\usepackage[skip=0.2cm]{caption}
\journal{Pharmacological Research}

\begin{document}
\begin{frontmatter}

\title{Deep Learning Prediction of Adverse Drug Reactions Using Open TG–GATEs and FAERS Databases}

\author[1]{Attayeb Mohsen\corref{cor1}}
\cortext[cor1]{Corresponding authors}
\ead{attayeb@nibiohn.go.jp}

\author[1]{Lokesh P. Tripathi}
\ead{lokesh@nibiohn.go.jp}

\author[1,2]{Kenji Mizuguchi\corref{cor1}}
\ead{kenji@nibiohn.go.jp}
\address[1]{Artificial Intelligence Center for Health and Biomedical Research (ArCHER), National Institutes of Biomedical Innovation, Health and Nutrition, Osaka, Japan. }
\address[2]{Institute for Protein Research, Osaka University, Osaka, Japan.}




\begin{abstract}
With the advancements in Artificial intelligence ({AI}) and the accumulation of health-related big data, it has become increasingly feasible and commonplace to leverage machine learning technologies to analyze clinical and omics metadata to assess the possibility of adverse drug reactions or events {(ADRs)} in the course of drug discovery. Here, we have described a novel approach that combined drug-induced gene expression profile from Open TG–GATEs (Toxicogenomics Project–Genomics Assisted Toxicity Evaluation Systems) and ADR occurrence information from FAERS (FDA [Food and Drug Administration] Adverse Events Reporting System) database to predict the likelihood of ADRs.

We generated a total of 14 models using Deep Neural Networks (DNN) to predict different ADRs; in the validation tests, our models achieved a mean accuracy of 85.71\%, indicating that our approach successfully and consistently predicted ADRs for a wide range of drugs. As an example, we have described the ADR model in the context of Duodenal ulcer. 

We believe that our models will help predict the likelihood of ADRs while testing novel pharmaceutical compounds, and will be useful for researchers in drug discovery.
\end{abstract}


\end{frontmatter}

%\linenumbers

\input{sections}

\section*{Declaration of Competing Interest}
The authors state that there is no conflict of interest

\section*{Acknowledgemnts}
We would like to show our gratitude to Dr. Yoshinobu Igarashi (Laboratory of Toxicogenomics Informatics, NIBIOHN) for his valuable suggestions and comments. However, the statements made here are solely the responsibility of the authors. No external funding was utilized in this study.

\bibliography{main}

\end{document}

%% file: sections.tex
%%%%%%%%%%%%%%%%%%%%%%%%%%%%%%%%%%%%%%%%%%
\section{Introduction}
%%%%%%%%%%%%%%%%%%%%%%%%%%%%%%%%%%%%%%%%%%
Adverse drug reaction (ADR) or event is defined as any unintended or undesired effect of a drug \cite{Coleman2016,Katzung2012}. ADRs are responsible for a high number of visits to emergency departments and in-hospital admissions. The Japan Adverse Drug Events (JADE) study reported around 17 adverse drug events per 1000 patient days. 1.6\% were fatal, 4.9\% were life-threatening, and 33\% were serious \cite{Morimoto2010}; these observations underscore the importance of toxicity assessment of any medication, especially in the early stages of drug discovery. 

\textit{In--silico} approaches to ADR prediction have been promising in terms of accuracy and underlying mechanisms understanding. Researchers have variously utilized multiple data types ranging from chemical structures to literature mining to predict ADRs \cite{Ho2016}.

Machine learning methods can play a significant role in the interpretation of various data types to predict ADRs. Deep learning \cite{Wang2020}, a subset of machine learning in Artificial intelligence (AI), has emerged as a promising and highly effective approach that can combine and interrogate diverse biological data types to generate new hypotheses. 
Deep learning methods are used extensively in the field of drug discovery and drug repurposing; however, their applications in ADR prediction are rather limited \cite{Dana2018,Vamathevan2019, Zhang2017}. 

Open TG–GATEs \cite{Igarashi2014} is a large--scale toxicogenomics database that collects gene expression profiles of \textit{in vivo} as well as \textit{in vitro} samples that have been treated with various drugs. 
These expression profiles are an outcome of the Japanese Toxicogenomics Project \cite{Uehara2009}, which aimed to build an extensive database of drug toxicities for drug discovery. It also collects physiological, biochemical, and pathological measurements of the treated animals. Similar databases that aim to profile compund toxicities have also been developed \cite{AlexanderDann2018, Chen2012}.

In contrast with other databases, such as (LINCS) \cite{Subramanian2017}, which have been used to predict multiple ADRs in a single study\cite{Wang2016}, {Open TG--GATEs} has been used to investigate individual/specific toxicities \cite{RuedaZrate2017}. To the best of our knowledge, no multiple ADR predictions have been attempted using Open TG--GATEs. 

The design of Open TG-Gates has several advantages over the {LINCS} database, chiefly the inclusion of \textit{in vivo} samples with different doses and durations of administration. Therefore, we designed our analysis to encompass multiple samples with different dosages and duration for each compound, necessitating additional noise-removal steps in the data set.

In this study, we describe our deep learning-based, systematic ADR prediction models that combine ADR occurrence data, including frequency details, from the FAERS (FDA Adverse Event Reporting System)  database, with the gene expression profiles from Open TG-GATEs. We aim to improve the models' performance by applying recently developed feature selection and hyper-parameter optimization algorithms. The methodologies and models described in our study offer useful tools for assessing the likelihood of ADRs and for incorporation into new drug development pipelines.

%%%%%%%%%%%%%%%%%%%%%%%%%%%%%%%%%%%%%%%%%%
\section{Materials and Methods}
%%%%%%%%%%%%%%%%%%%%%%%%%%%%%%%%%%%%%%%%%%

\subsection{Overview}

An overview of this study's methodology is shown in Figure:~\ref{fig:flowchart}. First, we retrieved the relevant data from the above-described two databases (open TG--GATEs and FAERS). We pre-processed the gene expression data to filter out noisy profiles by using a simple classification model, and retained only the significant ADR-drug associations (p<0.05; Fisher's exact test). 
Next, gene expression profile datasets were created by assigning positive and negative compounds for each ADR. We split the datasets into training and validation sets five times. We used the training set data to perform feature selection and build deep neural network models with hyper-parameter tuning using the Optuna package (see below). Finally, we evaluated the performances of the individual models on the validation set. We discuss these steps in detail below:

\begin{figure}
    \centering
    \includegraphics[width=0.5\hsize]{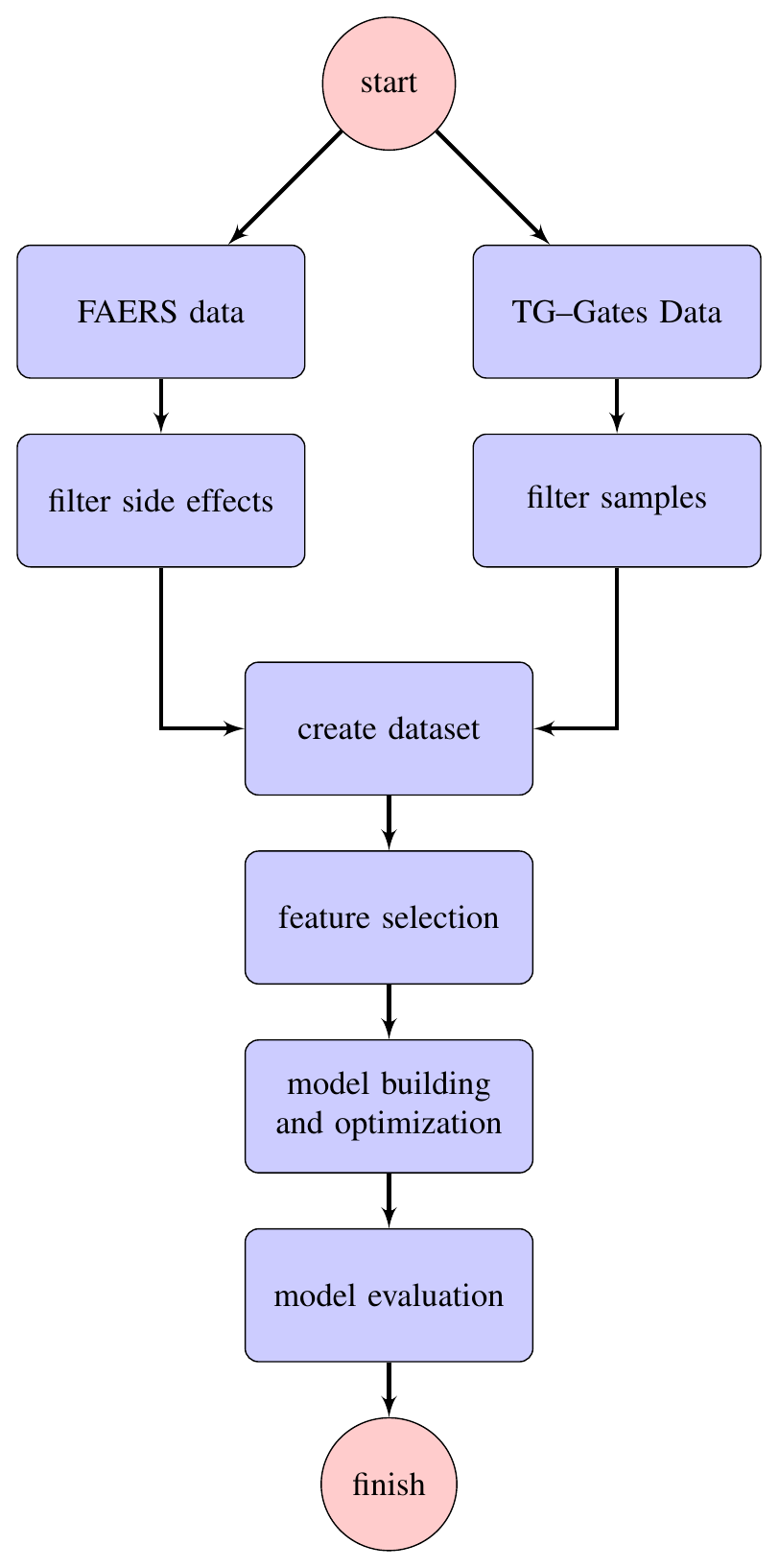}
    \caption{Overall analysis flowchart}
    \label{fig:flowchart}
\end{figure}

\subsection{Data retrieval and processing:} 
\subsubsection{Open TG--GATEs database:} 
%\paragraph{preprocessing: }
We extracted the \textit{in--vivo} gene expression profiles of rat liver samples from the Open TG--GATEs database \cite{Igarashi2014, Uehara2009}.
We selected the rat \textit{in-vivo} data  for our analysis chiefly because the \textit{in-vivo} dataset included more compounds and a greater number of time points as compared with the \textit{in-vitro} data (rat and human). However, our methodology can be easily extended to the other datasets.

This dataset was comprised of single-dose experiments and repeated-dose experiments. Single-dose experiments included injection-to-sacrifice periods of 3, 6, 9, or 24 hours, whereas, in the repeated dose experiments, rats were injected once daily for 4, 8, 15, or 29 days. In the repeated-dose experiments, all rats were sacrificed 24 hours after the last dose \cite{Igarashi2014}.
In Open TG--GATEs, gene expression profiles were measured using Microarray technology (Affymetrix GeneChip). 

The Affymetrix CEL files were downloaded from \url{http://toxico.nibiohn.go.jp}, and were preprocessed using the affy package \cite{Gautier2004} from R Bioconductor (\url{https://bioconductor.org/}); Affymetrix Microarray Suite algorithm version 5 (mas5) was applied with the default parameters provided in affy, wherein normalization = TRUE. The resulting normalized dataset -- hereafter referred to as ``the raw dataset'' -- was used for all the subsequent analyses. Next, the fold change values were calculated for each probe set by dividing the raw dataset by the mean intensities of corresponding control samples; these values were then log2 transformed, hereafter referred to as the ``log2FC dataset''.

Since the experiment design contained multiple dosages and durations of exposure, the drugs had varied effects on the gene expression profiles. To reduce the noise, we clustered all the samples to classify them as either treated or control using general linear modeling with Lasso penalty machine learning model (GLMNET package from R \cite{Friedman2010}.). We used the whole raw dataset as the training set with a two-class classification (treated and control). We fed through all the microarray data of the same duration to a single model, creating one model for each exposure set duration. Next, we estimated the probability of being classified as a treated sample for all the training sets. Only those samples with a probability of higher than 92\% were included in our analysis. The remaining samples were considered to fall within the gray zone between treated and control, and they were discarded. We chose a cut-off of 92\% because, at this threshold, no control samples were misclassified as treated.

\subsubsection{Standardized FAERS data:} 
FAERS (FDA Adverse Event Reporting System) is ``a database that collects adverse event reports, medication error reports, and product quality complaints resulting in adverse events that were submitted to FDA'' (\url{https://open.fda.gov/data/faers/}). Since the terms used in the FAERS database are left to the reporter to decide, inaccurate descriptions can often be incorporated, such as using general, vague terms to describe adverse events or treatments \cite{Wong2015}. To surmount this issue, we used the portion of the FAERS dataset standardized by Banda et al. \cite{Banda2016}. They had curated and standardized the entries of the FAERS database for 11 years (2004 to 2015) following Medical Dictionary for Regulatory Activities ({MedDRA}) terms \cite{Wood1994}. 

We extracted all the compound-ADR combinations (70,553,900) from the total number of reports (4.8 million). Among the difficulties of using the FAERS database in ADRs` prediction models is the presence of reports with multiple drugs used (Multipharma), which is expected in patients with chronic diseases. Such cases introduce unreliable associations added to the data noise. To solve this issue, we used only the association in which the drug is assigned as the primary suspect (PS) (15,377,900). We calculated the number of reports for each compound-adverse drug event combination and calculated the total number of reports of the compound in question and also the total number of reports of the adverse event. 

Using this data, we performed the Fisher test \cite{Ghosh1988} using ``fisher.test'' function from R with the parameter (alternative = ``greater''). This option returns a significant P-value only in the event of a positive association, in contrast to ``two.sides'' test, which assesses both positive and negative associations Table~\ref{tab:fisherexplain}.

\begin{table}[H]
    \caption{Fisher exact test: a: the number of reports of that the compound cause the ADE, b: the number of reports of the compound that does not report the cause of ADE, c: the number of all positive reports of the ADE for all compounds other than the specific compound, e: the number of all negative reports of all compound other than the specific compound. }
    \label{tab:fisherexplain}
    \centering
\begin{tabular}{|l|c|c|c|}
\toprule
 & Positive & Negative & Row total \\
\midrule
 compound & a & b & a+b \\
\midrule
all other compounds & c & d & c+d \\
\midrule
Column total & a+c & b+d & all reports \\
\bottomrule
\end{tabular}
\end{table}

\subsection{Model building and training}
% === Reviewer comment ===
% 2) Are the hyperparameters optimized with respect to the validation accuracy or to the training accuracy? Or do you employ, for each model, a train/validation/test split? If so, which is the splitting ratio, and how many samples are contained in the dataset? 
The training set included only treated samples (excluding all controls). For each model, we designated the compounds with the most significant associations as positive compounds, (p--value threshold $<$ 0.05) and the least significantly associated compounds as negative compounds. We evenly balanced the number of positive and negative compounds for each model.

When assembling the training and validation sets, we imposed two criteria: 1) the data-sets were balanced, i.e., the number of positive and negative samples were equal in both the sets and; 2) no compounds were commonly shared between training and validation. Instead of following a standardized cross-validation method, we created five models for each side effect by exploring all combinations of compounds in training and validation sets. However, we kept only those that comply with the previous conditions. 

%Feature selection
To prevent information leakage between the validation set and the training set, feature selection was performed using the training set data only, validation set data was not involved at all. Moreover, the hyper-parameter optimization was also done using the information from the training set only, and validation set accuracy was only used to identify the best model.

For feature selection, we used Boruta \cite{Kursa2010} implementation in Python \url{https://github.com/scikit-learn-contrib/boruta_py} , that is based on the random forest classifier from  scikit-learn \cite{Pedregosa2011} python  package with default parameters. Boruta algorithm utilizes the random forest classifier variable importance feature. To remove the effect of randomness and improve the accuracy of important feature detection, Boruta creates extra shadow variables by shuffling the values of the original features, then the feature importance is measured. Features with significantly higher importance than the shadow variables are considered important, while those with less importance are considered less important. This procedure was repeated 100 times to detect the important features more accurately.\cite{Kursa2010} 
% Deep NN
Then we used {TensorFlow 2}\cite{Abadi2016} to construct deep learning models. 
The input of the deep learning model is constructed of two dimensional matrix with samples in rows and genes on columns.

Each model consists of three groups of layers, input, output, and hidden layers. We applied Optuna \cite{Akiba2019} for hyperparameters tuning. Optuna uses the trial and error method for optimization, by randomly assigning values to the model hyperparameters from a range of values or choices offered by the author for a pre-determined number of trials. Subsequently, the results of all the trials can be examined to determine the most optimal parameters. 
The parameters that were optimized included: 
``depth'': corresponds to the number of densely connected layers (number of hidden layers aka DNN depth); 
the possible values are  [1, 2, 5, 10, 30]. 
``width'': corresponds to the number of nodes per layer, 
the possible values were [100, 250, 500, 700].
To reduce the likelihood of over fitting, we used two measures. The first measure
``drop'' is to drop some nodes before going to the next layer. It took one of these values [0.2, 0.3, 0.4, 0.5], (0.2 means 20\% of nodes are dropped).
The second measure is the introduction of noise: 
the value of introduced gaussian noise [0.2, 0.3, 0.4, 0.5]. Other hyperparameters included: 
``activation'': corresponds to the activation function of the final layer (output layer); possible values: ['sigmoid', 'linear'], 
``learning rate'' for Adam optimizer \cite{Diederik2014} was selected among [0.001, 0.0005, 0.00001]. 

The models with highest validation set accuracy were chosen.
\begin{table}[H]
    \caption{Optuna hyper--parameter choices}
    \label{tab:optunaexplain}
    \centering
\begin{tabular}{l l}
\toprule
 Parameter & choices \\
\midrule
 depth & 1, 2, 5, 10, 30 \\
 width & 100, 250, 500, 700 \\
 drop percentage & 0.2, 0.3, 0.4, 0.7 \\
 gaussian noise & 0.2, 0.3, 0.4, 0.5 \\
 activation & sigmoid, linear \\
 learning rate & 0.001, 0.0005, 0.00001 \\
\bottomrule
\end{tabular}
\end{table}

The maximum number of epochs was set at 800; however, we applied the ``early stopping'' strategy if the accuracy did not improve for 75 epochs. The best models were saved for each Optuna trial. The best parameters are shown in the Supplementary Table: S1.

\begin{figure}
    \centering
    \includegraphics[width=\hsize]{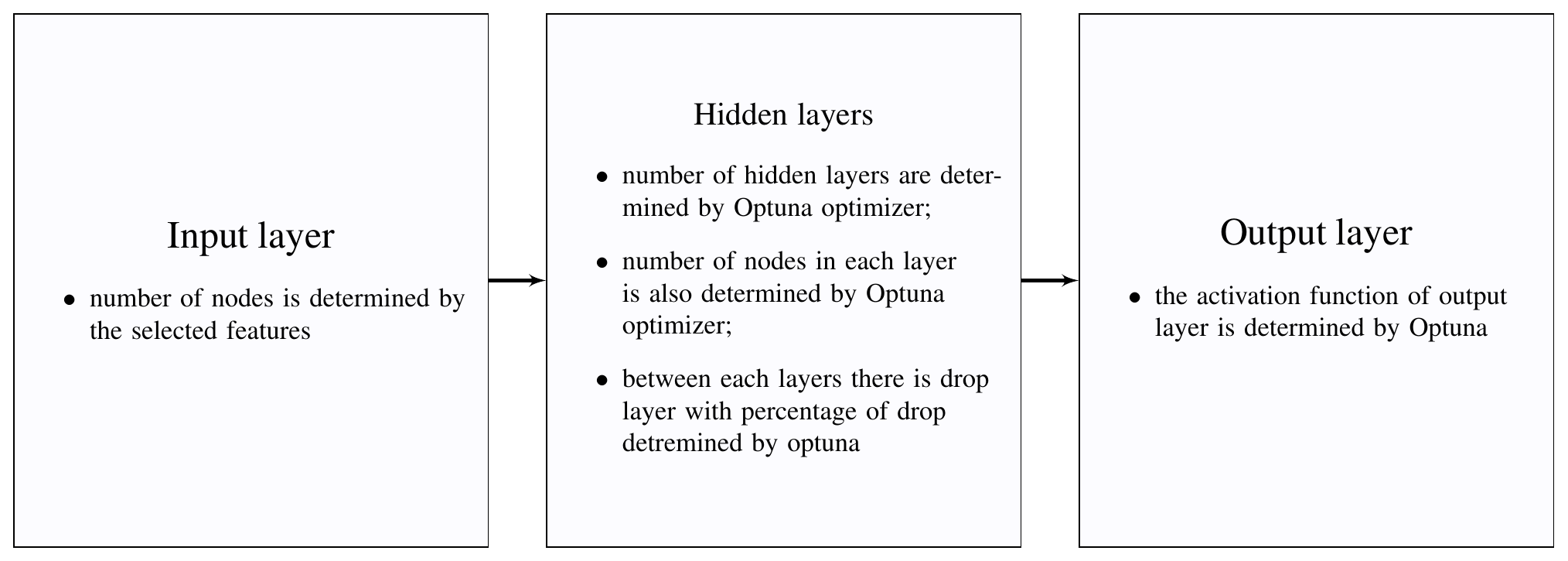}
    \caption{DNN structures}
    \label{fig:dnnstructure}
\end{figure}

\subsection{Evaluation and enrichment analysis}
Model performances were evaluated by testing the performance of the validation set prediction. We estimated the accuracy of the validation set and the area under the ROC (Receiver operating characteristic) curve using the scikit-learn \cite{Pedregosa2011} package from Python.

TargetMine data analysis platform was used for enrichment analysis and gene annotation\cite{Chen2019}. Databases used for enrichment analysis were KEGG, Reactome, and NCI. \textit{P--values} were calculated in TargetMine using one-tailed Fisher's exact test. Multiple test correction was set to Benjamini Hochberg, with \textit{p-value} significance threshold of 0.05.

%%%%%%%%%%%%%%%%%%%%%%%%%%%%%%%%%%%%%%%%%%
\section{Results}
%%%%%%%%%%%%%%%%%%%%%%%%%%%%%%%%%%%%%%%%%%
\subsection{Data processing}
To reduce the data dispersion caused by multiple dose levels and injection durations (sacrifice period) in Open TG-GATEs, we filtered out low quality/unsuitable samples. To do that, we used Lasso to classify the samples to either treated or control classes. A total of 6619 of 10573 treated samples, chiefly belonging to the ``Low'' dose level category, were classified as controls and eventually excluded. Samples that were correctly classified as treated (3953 samples) remained for subsequent analysis. (Table~\ref{tab:includedsamplesinformation}).

\begin{table}[H]
    \caption{The number of Open TG-GATEs samples included in the analysis after clustering using Lasso, dose level and sacrifice period details are shown.}
    \label{tab:includedsamplesinformation}
    \centering
\begin{tabular}{llrr}
\toprule
                 &        &  Original &  Included \\
\midrule
Dose Level       & Low    &      3540 &       421 \\
                 & Middle &      3537 &      1212 \\
                 & High   &      3496 &      2320 \\
\midrule
Sacrifice Period & 24 hr  &      1408 &       762 \\
                 & 9 hr   &      1371 &       556 \\
                 & 6 hr   &      1371 &       533 \\
                 & 3 hr   &      1368 &       518 \\
                 & 4 day  &      1275 &       254 \\
                 & 8 day  &      1275 &       472 \\
                 & 15 day &      1266 &       500 \\
                 & 29 day &      1239 &       358 \\
\bottomrule
\end{tabular}
\end{table}

\subsection{Model building and training}
We created a total of 14 models (Table:~\ref{tab:createdmodels}). The number of compounds used to create each model ranged from 10 to 18. We equalized the number of positive compounds and negative compounds to generate balanced models.

\begin{table}[H]
    \caption{The number of compounds, and samples used to create ADRs prediction models. (AGEP: Acute generalized exanthematous pustulosis, ECG: Electrocardiogram). (+): Positive , (-): Negative}
    \label{tab:createdmodels}
    \centering
\begin{tabular}{lcccc}
\toprule
      ADR & \multicolumn{2}{c}{Drugs} & \multicolumn{2}{c}{Samples} \\
      & + & - & + & - \\
\midrule
 AGEP &  5 &  5 &  203 &  87 \\
 Bone marrow failure &  5 &  5 &  125 &  178 \\
 Catatonia &  5 &  5 &  169 &  130 \\
 Duodenal ulcer &  6 &  6 &  122 &  178 \\
 ECG qt prolonged &  5 &  5 &  105 &  152 \\
 Febrile neutropenia &  5 &  5 &  62 &  141 \\
 Gastric haemorrhage &  5 &  5 &  124 &  161 \\
 Hepatitis fulminant &  5 &  5 &  193 &  126 \\
 Liver transplant &  9 &  9 &  297 &  244 \\
 Lymphocytosis &  5 &  5 &  196 &  128 \\
 Neutropenic sepsis &  6 &  6 &  89 &  165 \\
 Optic atrophy &  6 &  6 &  122 &  139 \\
 Torsade de pointes &  7 &  7 &  144 &  219 \\
 Toxic epidermal necrolysis &  5 &  5 &  136 &  99 \\
\bottomrule
\end{tabular}
\end{table}

\subsection{Model evaluation}
\begin{figure}
    \centering
    \includegraphics[width=\hsize]{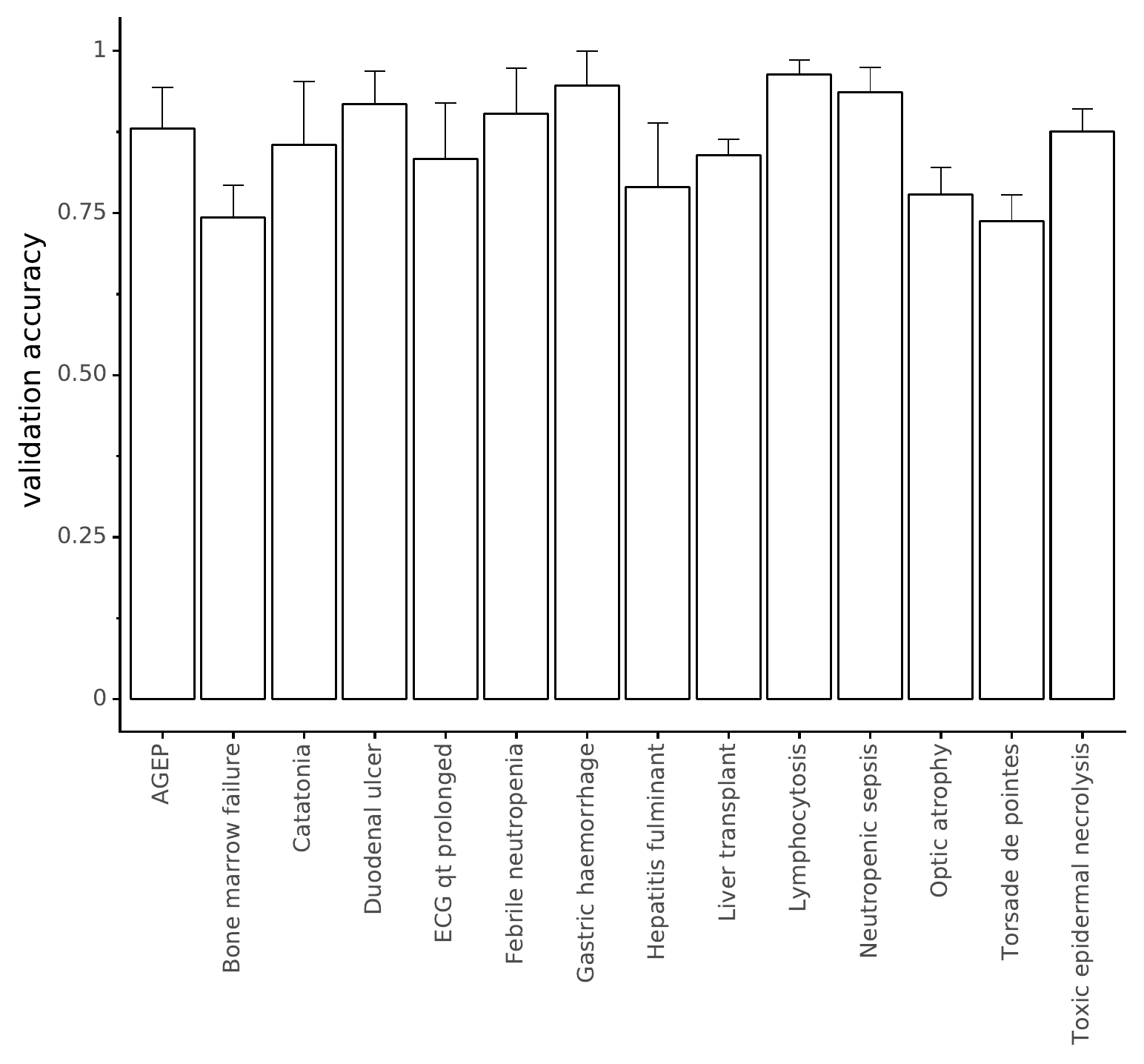}
    \caption{Validation accuracy of the created models}
    \label{fig:validationaccuracy}
\end{figure}

\begin{figure}
    \centering
    \includegraphics[width=\hsize]{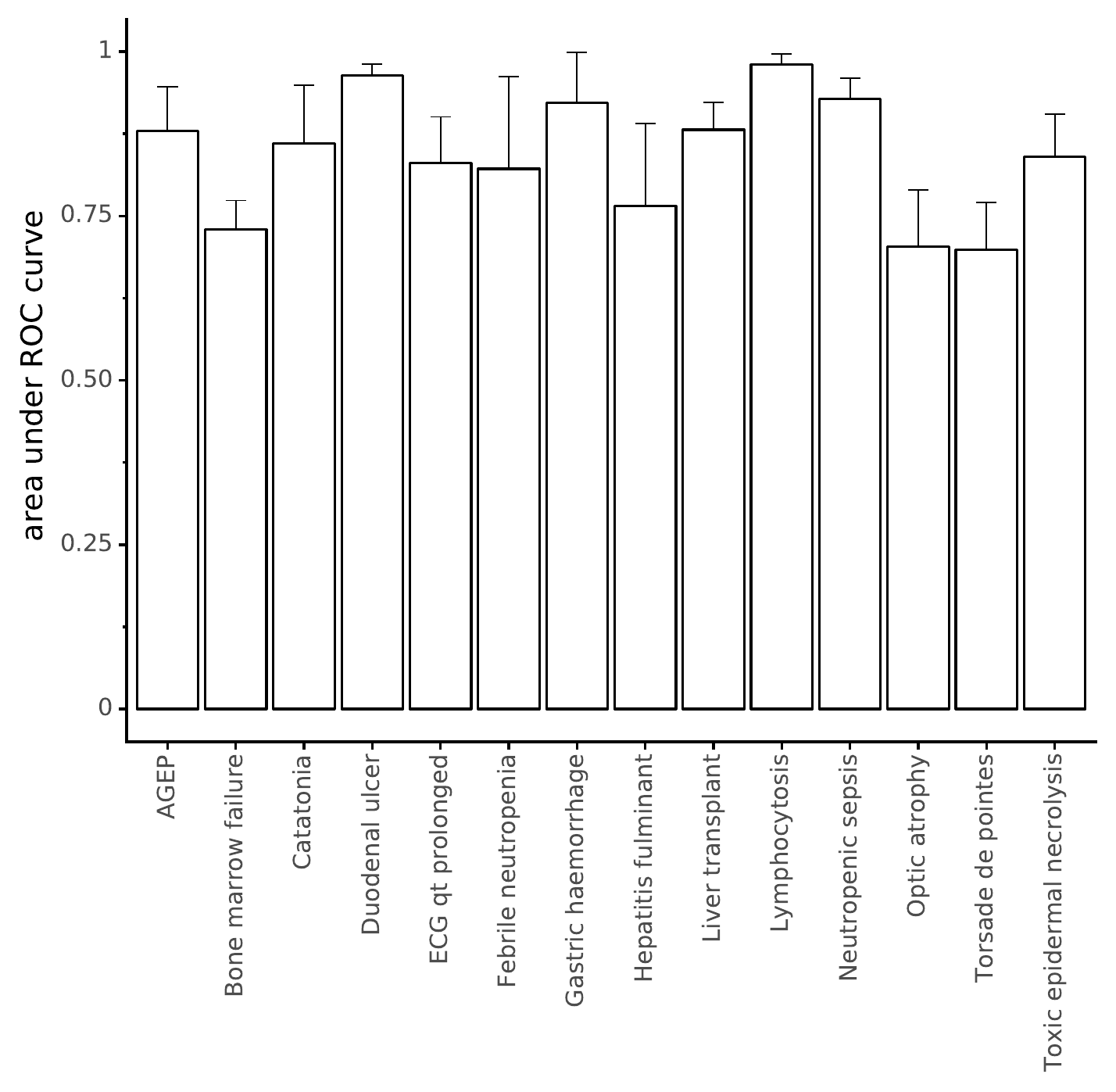}
    \caption{Area under ROC curves for the created models.}
    \label{fig:auc}
\end{figure}

The Average accuracy for all models was 85.71\% [minimum$=$67.9\%, and maximum=100\%, standard error=0.1]. The validation accuracies of the models are shown in Figure~\ref{fig:validationaccuracy}. The area under the Receiver Operating Characteristic (ROC) curve is shown in Figure:~\ref{fig:auc}.

\subsection{Case study: Duodenal Ulcer}
To highlight the effectiveness of our approach, we describe below our observations on the development of the duodenal ulcer ADR prediction model.
Duodenal ulcer is a type of peptic ulcer disease characterized by the emergence of open sores on the duodenum's inner lining of the duodenum \cite{Kuna2019}. It is mainly caused by the failure of gastrointestinal system inner coating protection, and the most common causing agents are \textit{Helicobacter Pylori} infection and NSAIDs (Non-steroidal anti-inflammatory drugs). It can lead to serious bleeding or perforation. 

\subsubsection{Duodenal ulcer model description}
\begin{figure}
    \centering
    \includegraphics[width=\hsize]{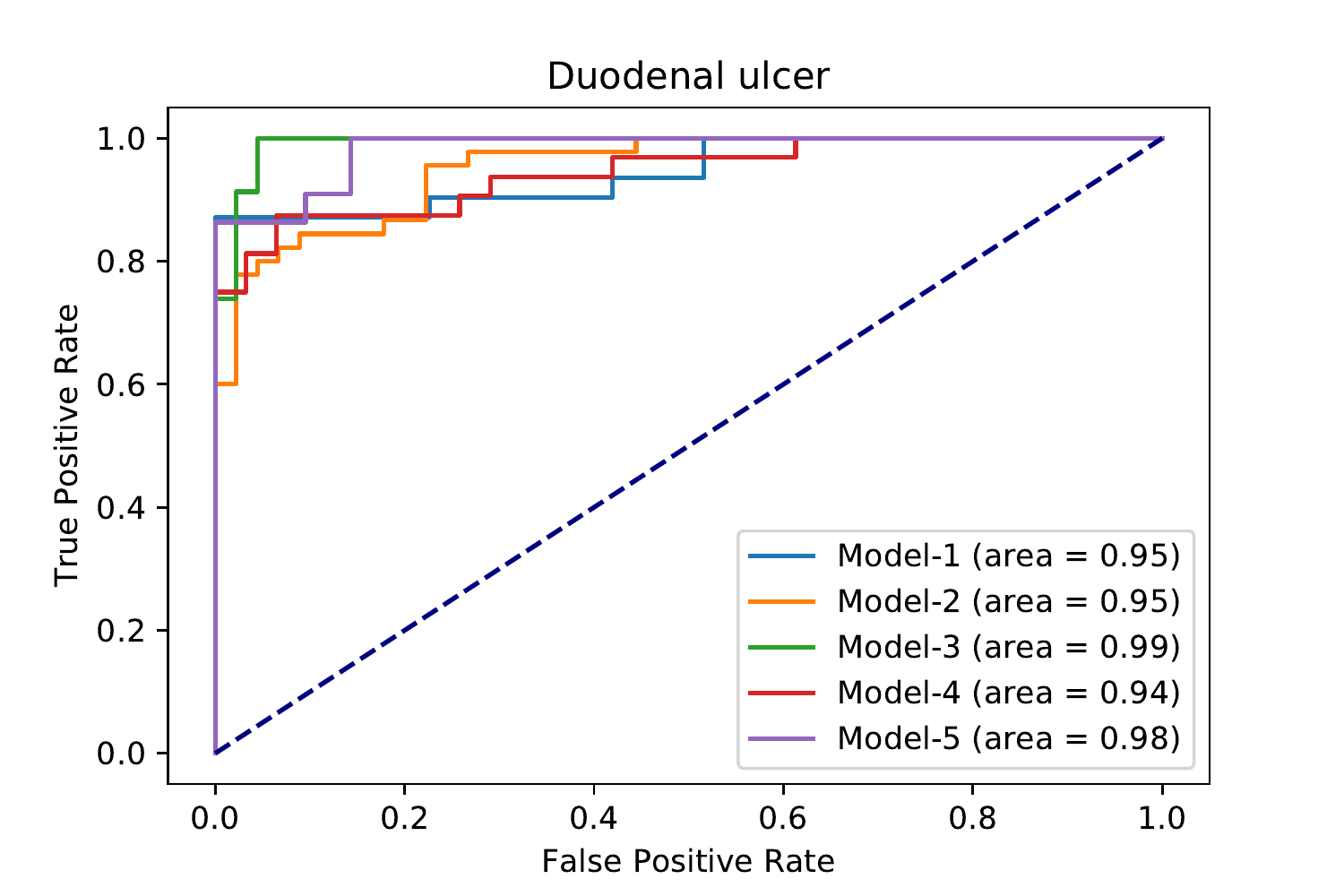}
    \caption{Area under ROC curves for duodenal ulcer models. Each color corresponds to a different model.}
    \label{fig:duroc}
\end{figure}

Using the FAERS database, the six most significantly associated drugs with duodenal ulcer were identified using Fisher test (positive drugs: Aspirin, Diclofenac, Ibuprofen, Indomethacin, Meloxicam, Naproxen) and the least associated drugs were also identified (negative drugs: Acetaminophen, Amiodarone, Bortezomib, Carbamazepine, Ciprofloxacin, Cyclophosphamide). Subsequently, the open TG--Gates samples of these drugs were used to build the prediction models.
Details of the compounds used, the number of the gene expression samples associated with these compounds, and the results of the Fisher test (see Methods) are shown in Table~\ref{tab:dudataset}. 
The ROC curves and the area under them for the five models of duodenal ulcer (each trained on a different training set) are shown in Figure~\ref{fig:duroc}. The performance of these models showed that the area under the curve ranges from 0.94 to 0.99. 
The number of the features (genes) commonly selected among the five duodenal ulcer models were 108.

\begin{table}
    \caption{Details of the data set for the duodenal ulcer model (compounds, the number of the samples, the p-value of the Fisher test and the class in the training or test sets). Positive class compounds are those that can cause doudenal ulcer, while Negative class compounds are controls. Entries were ordered alphabetically}
    \label{tab:dudataset}
    \centering
%\begin{tabular}{lrrl}
\begin{tabular}{lrrl}
\toprule
{} &  Fisher (\textit{p}) &  No &     Class \\
\midrule
Acetaminophen    &  \SI{4.88e-01} &  50 &  negative \\
Amiodarone       &  \SI{9.35e-01} &  31 &  negative \\
Aspirin          &  \SI{4.67e-233} &  45 &  positive \\
Bortezomib       &  \SI{3.57e-01} &  24 &  negative \\
Carbamazepine    &  \SI{9.99e-01} &  45 &  negative \\
Ciprofloxacin    &  \SI{9.67e-01} &  7 &  negative \\
Cyclophosphamide &  \SI{8.29e-01} &  21 &  negative \\
Diclofenac       &  \SI{1.3e-66} &  9 &  positive \\
Ibuprofen        &  \SI{1.22e-112} &  24 &  positive \\
Indomethacin     &  \SI{5.40e-15} &  12 &  positive \\
Meloxicam        &  \SI{2.15e-35} &  10 &  positive \\
Naproxen         &  \SI{2.02e-78} &  22 &  positive \\
\bottomrule
\end{tabular}

\end{table}
%}

\subsubsection{Enrichment Analysis}
Pathway enrichment analysis (see Methods) using the duodenal ulcer-selected features (Table: S2) clearly highlighted the involvement of bleeding cascade and complement function (Table:~\ref{tab:enrichmentanalysis}).

The manifestation of a duodenal ulcer activates the complement cascade, which is probably due to the inflammation caused by the acid effect on the intestinal mucosa. Bleeding is also linked with the duodenal ulcer disease. The enrichment of the Fatty acid degradation pathway is consistent with the fact that the majority of the compounds that cause duodenal ulcers are NSAIDs that inhibit Arachidonic acid metabolism, which is a part of the Fatty acid metabolism pathway.

\begin{table}[H]
    \caption{The enrichment analysis results of duodenal ulcer model, showing the involvement of both complement and coagulation functions}
    \label{tab:enrichmentanalysis}
    \centering
\begin{tabular}{p{5cm}lp{0.01cm}}
\toprule
Pathway &  \textit{P}value & \\
\midrule
Complement and coagulation cascades [rno04610] & \SI{1.86e-11} & \\
Regulation of Complement cascade [R-RNO-977606] & \SI{1.56e-5} &\\
Complement cascade [R-RNO-166658] & \SI{9.16e-5} &\\
Pertussis (rno05133) & \SI{3.84e-3} & \\
Fatty acid degradation (rno00071) & \SI{4.83e-3} &\\
\bottomrule
\end{tabular}
\end{table}

%%%%%%%%%%%%%%%%%%%%%%%%%%%%%%%%%%%%%%%%%%
\section{Discussion}
%%%%%%%%%%%%%%%%%%%%%%%%%%%%%%%%%%%%%%%%%%

We have described a novel approach that combined toxicogenomics gene expression profiles extracted from Open TG-GATEs and ADRs reports extracted from FAERS to predict the likelihood of ADRs. 
This integration of two highly distinct data types allowed us to predict ADRs successfully. Moreover, it led to creating a novel dataset that associated drug-induced gene expression profiles with ADRs.

To overcome the significant challenges in combining the two datasets, we first sought to extract the individual drug-induced gene expression signature from Open TG-GATEs. Next, we extracted the ADR occurrence frequencies for these drugs and estimated their statistical significance to eventually combine the two datasets.

Moreover, due to multiple dose-levels and sacrifice periods and the presence of repeated and single injection events, the drug-induced gene expression profiles were fairly noisy. We generated a simple model to classify all the samples as either control or treated classes using  Lasso to filter out this noise. We performed a rigorous statistical assessment to narrow down suitable samples for subsequent analyses (see Methods for details).

Recently, deep learning has gathered an increasing usage in the field of drug discovery \cite{Zhang2017, Lee2019}. In this study we have used deep learning together with feature selection to reduce the data dimensionality and avoid overfitting due to limited samples.

Previously,  Wang et al. \cite{Wang2016} utilized multiple cell lines to develop predictive models for multiple ADRs. Another study \cite{Joseph2013} demonstrated that blood transcriptomics could be used to examine other organ toxicities. Our results have supported this notion by exhibiting robust prediction models with high accuracy using liver samples. Liver is a vital organ for drug metabolism and receives a significant amount of blood, and hence, it is widely used in drug toxicity studies. Moreover, even in the absence of pathological responses to the compound toxicity, cells still display differences in gene expression profiles.

We did not apply the leave-p-out or k-fold cross-validation protocols to the gene expression samples because: 1) the training and validation sets should be assigned based on compounds segregation, i.e., the same compound should not span training and validation sets, and 2) the number of samples differed from compound to compound and thus, creating balanced sets was impossible. We instead adopted the approach of creating five models with different training and validation sets. We also performed feature selection for each of the set combinations.

This study utilized \textit{in vivo} gene expression data in contrast with another study \cite{Wang2016} that utilized the data from the {LINCS} database \cite{Subramanian2017}, a collection of \textit{in vitro} gene expression profiles from human cell lines. Our approach is easily applicable to other publicly available collections of toxicogenomics data, such as those from Drug Matrix \cite{drugmatrix}. Another difference is that Wang \textit{et al.} \cite{Wang2016} combined chemical structure and Gene Ontology (GO) term associations in their models. They also selected only a single gene expression profile to represent the compound effects; in contrast, our method analyzed multiple samples with different doses and durations of each compound's exposure. Hence, our method is better equipped to account for the biological variations that are inherent in drug-induced physiological and phenotypic responses. Indeed, our models performed better than Wang \textit{et al.} 's gene expression data-only models \cite{Wang2016}. This difference in performances may probably be attributed to our utilization of multiple samples for each compound.

Using the Optuna optimization package \cite{Akiba2019} made these prediction models' creation computationally expensive; hence, only a limited number of models were built. However, our approach can help generate models to serve specific applications using other data resources such as DrugMatrix, depending on the user's needs.

In conclusion, we have developed 14 Deep learning models to predict adverse drug events utilizing the public available Open TG–Gates and FAERS databases. These models can be used to examine if a new drug candidate can cause these side effects. Moreover, following the same feature selection and model building and tuning steps, other models can be created for other ADRs.

%% file: main.bbl
\begin{thebibliography}{10}
\expandafter\ifx\csname url\endcsname\relax
  \def\url#1{\texttt{#1}}\fi
\expandafter\ifx\csname urlprefix\endcsname\relax\def\urlprefix{URL }\fi
\expandafter\ifx\csname href\endcsname\relax
  \def\href#1#2{#2} \def\path#1{#1}\fi

\bibitem{Coleman2016}
J.~J. Coleman, S.~K. Pontefract,
  \href{https://doi.org/10.7861/clinmedicine.16-5-481}{Adverse drug reactions},
  Clinical Medicine 16~(5) (2016) 481--485.
\newblock \href {https://doi.org/10.7861/clinmedicine.16-5-481}
  {\path{doi:10.7861/clinmedicine.16-5-481}}.
\newline\urlprefix\url{https://doi.org/10.7861/clinmedicine.16-5-481}

\bibitem{Katzung2012}
B.~G. Katzung, Development \& regulation of drugs, in: B.~G. Katzung, S.~B.
  Masters, A.~J. Trevor (Eds.), Basic \& Clinical Pharmacology, McGraw Hill,
  2012, Ch.~5, pp. 69--78.

\bibitem{Morimoto2010}
T.~Morimoto, M.~Sakuma, K.~Matsui, N.~Kuramoto, J.~Toshiro, J.~Murakami,
  T.~Fukui, M.~Saito, A.~Hiraide, D.~W. Bates,
  \href{https://doi.org/10.1007/s11606-010-1518-3}{Incidence of adverse drug
  events and medication errors in japan: the {JADE} study}, Journal of General
  Internal Medicine 26~(2) (2010) 148--153.
\newblock \href {https://doi.org/10.1007/s11606-010-1518-3}
  {\path{doi:10.1007/s11606-010-1518-3}}.
\newline\urlprefix\url{https://doi.org/10.1007/s11606-010-1518-3}

\bibitem{Ho2016}
T.-B. Ho, L.~Le, D.~T. Thai, S.~Taewijit,
  \href{https://doi.org/10.2174/1381612822666160509125047}{Data-driven approach
  to detect and predict adverse drug reactions}, Current Pharmaceutical Design
  22~(23) (2016) 3498--3526.
\newblock \href {https://doi.org/10.2174/1381612822666160509125047}
  {\path{doi:10.2174/1381612822666160509125047}}.
\newline\urlprefix\url{https://doi.org/10.2174/1381612822666160509125047}

\bibitem{Wang2020}
X.~Wang, Y.~Zhao, F.~Pourpanah,
  \href{https://doi.org/10.1007/s13042-020-01096-5}{Recent advances in deep
  learning}, International Journal of Machine Learning and Cybernetics 11~(4)
  (2020) 747--750.
\newblock \href {https://doi.org/10.1007/s13042-020-01096-5}
  {\path{doi:10.1007/s13042-020-01096-5}}.
\newline\urlprefix\url{https://doi.org/10.1007/s13042-020-01096-5}

\bibitem{Dana2018}
D.~Dana, S.~Gadhiya, L.~S. Surin, D.~Li, F.~Naaz, Q.~Ali, L.~Paka, M.~Yamin,
  M.~Narayan, I.~Goldberg, P.~Narayan,
  \href{https://doi.org/10.3390/molecules23092384}{Deep learning in drug
  discovery and medicine; scratching the surface}, Molecules 23~(9) (2018)
  2384.
\newblock \href {https://doi.org/10.3390/molecules23092384}
  {\path{doi:10.3390/molecules23092384}}.
\newline\urlprefix\url{https://doi.org/10.3390/molecules23092384}

\bibitem{Vamathevan2019}
J.~Vamathevan, D.~Clark, P.~Czodrowski, I.~Dunham, E.~Ferran, G.~Lee, B.~Li,
  A.~Madabhushi, P.~Shah, M.~Spitzer, S.~Zhao,
  \href{https://doi.org/10.1038/s41573-019-0024-5}{Applications of machine
  learning in drug discovery and development}, Nature Reviews Drug Discovery
  18~(6) (2019) 463--477.
\newblock \href {https://doi.org/10.1038/s41573-019-0024-5}
  {\path{doi:10.1038/s41573-019-0024-5}}.
\newline\urlprefix\url{https://doi.org/10.1038/s41573-019-0024-5}

\bibitem{Zhang2017}
L.~Zhang, J.~Tan, D.~Han, H.~Zhu,
  \href{https://doi.org/10.1016/j.drudis.2017.08.010}{From machine learning to
  deep learning: progress in machine intelligence for rational drug discovery},
  Drug Discovery Today 22~(11) (2017) 1680--1685.
\newblock \href {https://doi.org/10.1016/j.drudis.2017.08.010}
  {\path{doi:10.1016/j.drudis.2017.08.010}}.
\newline\urlprefix\url{https://doi.org/10.1016/j.drudis.2017.08.010}

\bibitem{Igarashi2014}
Y.~Igarashi, N.~Nakatsu, T.~Yamashita, A.~Ono, Y.~Ohno, T.~Urushidani,
  H.~Yamada, \href{https://doi.org/10.1093/nar/gku955}{Open {TG}-{GATEs}: a
  large-scale toxicogenomics database}, Nucleic Acids Research 43~(D1) (2014)
  D921--D927.
\newblock \href {https://doi.org/10.1093/nar/gku955}
  {\path{doi:10.1093/nar/gku955}}.
\newline\urlprefix\url{https://doi.org/10.1093/nar/gku955}

\bibitem{Uehara2009}
T.~Uehara, A.~Ono, T.~Maruyama, I.~Kato, H.~Yamada, Y.~Ohno, T.~Urushidani,
  \href{https://doi.org/10.1002/mnfr.200900169}{The japanese toxicogenomics
  project: Application of toxicogenomics}, Molecular Nutrition {\&} Food
  Research 54~(2) (2009) 218--227.
\newblock \href {https://doi.org/10.1002/mnfr.200900169}
  {\path{doi:10.1002/mnfr.200900169}}.
\newline\urlprefix\url{https://doi.org/10.1002/mnfr.200900169}

\bibitem{AlexanderDann2018}
B.~Alexander-Dann, L.~L. Pruteanu, E.~Oerton, N.~Sharma, I.~Berindan-Neagoe,
  D.~M{\'{o}}dos, A.~Bender,
  \href{https://doi.org/10.1039/c8mo00042e}{Developments in toxicogenomics:
  understanding and predicting compound-induced toxicity from gene expression
  data}, Molecular Omics 14~(4) (2018) 218--236.
\newblock \href {https://doi.org/10.1039/c8mo00042e}
  {\path{doi:10.1039/c8mo00042e}}.
\newline\urlprefix\url{https://doi.org/10.1039/c8mo00042e}

\bibitem{Chen2012}
M.~Chen, M.~Zhang, J.~Borlak, W.~Tong,
  \href{https://doi.org/10.1093/toxsci/kfs223}{A decade of toxicogenomic
  research and its contribution to toxicological science}, Toxicological
  Sciences 130~(2) (2012) 217--228.
\newblock \href {https://doi.org/10.1093/toxsci/kfs223}
  {\path{doi:10.1093/toxsci/kfs223}}.
\newline\urlprefix\url{https://doi.org/10.1093/toxsci/kfs223}

\bibitem{Subramanian2017}
A.~Subramanian, R.~Narayan, S.~M. Corsello, D.~D. Peck, T.~E. Natoli, X.~Lu,
  J.~Gould, J.~F. Davis, A.~A. Tubelli, J.~K. Asiedu, D.~L. Lahr, J.~E.
  Hirschman, Z.~Liu, M.~Donahue, B.~Julian, M.~Khan, D.~Wadden, I.~C. Smith,
  D.~Lam, A.~Liberzon, C.~Toder, M.~Bagul, M.~Orzechowski, O.~M. Enache,
  F.~Piccioni, S.~A. Johnson, N.~J. Lyons, A.~H. Berger, A.~F. Shamji, A.~N.
  Brooks, A.~Vrcic, C.~Flynn, J.~Rosains, D.~Y. Takeda, R.~Hu, D.~Davison,
  J.~Lamb, K.~Ardlie, L.~Hogstrom, P.~Greenside, N.~S. Gray, P.~A. Clemons,
  S.~Silver, X.~Wu, W.-N. Zhao, W.~Read-Button, X.~Wu, S.~J. Haggarty, L.~V.
  Ronco, J.~S. Boehm, S.~L. Schreiber, J.~G. Doench, J.~A. Bittker, D.~E. Root,
  B.~Wong, T.~R. Golub, \href{https://doi.org/10.1016/j.cell.2017.10.049}{A
  next generation connectivity map: L1000 platform and the first 1, 000, 000
  profiles}, Cell 171~(6) (2017) 1437--1452.e17.
\newblock \href {https://doi.org/10.1016/j.cell.2017.10.049}
  {\path{doi:10.1016/j.cell.2017.10.049}}.
\newline\urlprefix\url{https://doi.org/10.1016/j.cell.2017.10.049}

\bibitem{Wang2016}
Z.~Wang, N.~R. Clark, A.~Ma'ayan,
  \href{https://doi.org/10.1093/bioinformatics/btw168}{Drug-induced adverse
  events prediction with the {LINCS} l1000 data}, Bioinformatics 32~(15) (2016)
  2338--2345.
\newblock \href {https://doi.org/10.1093/bioinformatics/btw168}
  {\path{doi:10.1093/bioinformatics/btw168}}.
\newline\urlprefix\url{https://doi.org/10.1093/bioinformatics/btw168}

\bibitem{RuedaZrate2017}
H.~A. Rueda-Z{\'{a}}rate, I.~Imaz-Rosshandler, R.~A. C{\'{a}}rdenas-Ovando,
  J.~E. Castillo-Fern{\'{a}}ndez, J.~Noguez-Monroy, C.~Rangel-Escare{\~{n}}o,
  \href{https://doi.org/10.1371/journal.pone.0176284}{A computational
  toxicogenomics approach identifies a list of highly hepatotoxic compounds
  from a large microarray database}, {PLOS} {ONE} 12~(4) (2017) e0176284.
\newblock \href {https://doi.org/10.1371/journal.pone.0176284}
  {\path{doi:10.1371/journal.pone.0176284}}.
\newline\urlprefix\url{https://doi.org/10.1371/journal.pone.0176284}

\bibitem{Gautier2004}
L.~Gautier, L.~Cope, B.~M. Bolstad, R.~A. Irizarry,
  \href{https://doi.org/10.1093/bioinformatics/btg405}{affy--analysis of
  affymetrix {GeneChip} data at the probe level}, Bioinformatics 20~(3) (2004)
  307--315.
\newblock \href {https://doi.org/10.1093/bioinformatics/btg405}
  {\path{doi:10.1093/bioinformatics/btg405}}.
\newline\urlprefix\url{https://doi.org/10.1093/bioinformatics/btg405}

\bibitem{Friedman2010}
J.~Friedman, T.~Hastie, R.~Tibshirani,
  \href{https://doi.org/10.18637/jss.v033.i01}{Regularization paths for
  generalized linear models via coordinate descent}, Journal of Statistical
  Software 33~(1) (2010).
\newblock \href {https://doi.org/10.18637/jss.v033.i01}
  {\path{doi:10.18637/jss.v033.i01}}.
\newline\urlprefix\url{https://doi.org/10.18637/jss.v033.i01}

\bibitem{Wong2015}
C.~K. Wong, S.~S. Ho, B.~Saini, D.~E. Hibbs, R.~A. Fois,
  \href{https://doi.org/10.1002/pds.3805}{Standardisation of the {FAERS}
  database: a systematic approach to manually recoding drug name variants},
  Pharmacoepidemiology and Drug Safety 24~(7) (2015) 731--737.
\newblock \href {https://doi.org/10.1002/pds.3805}
  {\path{doi:10.1002/pds.3805}}.
\newline\urlprefix\url{https://doi.org/10.1002/pds.3805}

\bibitem{Banda2016}
J.~M. Banda, L.~Evans, R.~S. Vanguri, N.~P. Tatonetti, P.~B. Ryan, N.~H. Shah,
  \href{https://doi.org/10.1038/sdata.2016.26}{A curated and standardized
  adverse drug event resource to accelerate drug safety research}, Scientific
  Data 3~(1) (5 2016).
\newblock \href {https://doi.org/10.1038/sdata.2016.26}
  {\path{doi:10.1038/sdata.2016.26}}.
\newline\urlprefix\url{https://doi.org/10.1038/sdata.2016.26}

\bibitem{Wood1994}
K.~L.~W. and, \href{https://doi.org/10.1002/pds.2630030105}{The medical
  dictionary for drug regulatory affairs ({MEDDRA}) project},
  Pharmacoepidemiology {\&} Drug Safety 3~(1) (1994) 7--13.
\newblock \href {https://doi.org/10.1002/pds.2630030105}
  {\path{doi:10.1002/pds.2630030105}}.
\newline\urlprefix\url{https://doi.org/10.1002/pds.2630030105}

\bibitem{Ghosh1988}
J.~K. Ghosh, \href{https://doi.org/10.1007/978-1-4612-3894-2_18}{A discussion
  on the fisher exact test}, in: Statistical Information and Likelihood,
  Springer New York, 1988, pp. 321--324.
\newblock \href {https://doi.org/10.1007/978-1-4612-3894-2_18}
  {\path{doi:10.1007/978-1-4612-3894-2_18}}.
\newline\urlprefix\url{https://doi.org/10.1007/978-1-4612-3894-2_18}

\bibitem{Kursa2010}
M.~B. Kursa, W.~R. Rudnicki,
  \href{https://doi.org/10.18637/jss.v036.i11}{Feature selection with {the
  Boruta Package}}, Journal of Statistical Software 36~(11) (2010).
\newblock \href {https://doi.org/10.18637/jss.v036.i11}
  {\path{doi:10.18637/jss.v036.i11}}.
\newline\urlprefix\url{https://doi.org/10.18637/jss.v036.i11}

\bibitem{Pedregosa2011}
F.~Pedregosa, G.~Varoquaux, A.~Gramfort, V.~Michel, B.~Thirion, O.~Grisel,
  M.~Blondel, P.~Prettenhofer, R.~Weiss, V.~Dubourg, J.~Vanderplas, A.~Passos,
  D.~Cournapeau, M.~Brucher, M.~Perrot, {{\'E}}douard Duchesnay,
  \href{http://jmlr.org/papers/v12/pedregosa11a.html}{Scikit-learn: Machine
  learning in python}, Journal of Machine Learning Research 12~(85) (2011)
  2825--2830.
\newline\urlprefix\url{http://jmlr.org/papers/v12/pedregosa11a.html}

\bibitem{Abadi2016}
M.~Abadi, A.~Agarwal, P.~Barham, E.~Brevdo, Z.~Chen, C.~Citro, G.~S. Corrado,
  A.~Davis, J.~Dean, M.~Devin, S.~Ghemawat, I.~Goodfellow, A.~Harp, G.~Irving,
  M.~Isard, Y.~Jia, R.~Jozefowicz, L.~Kaiser, M.~Kudlur, J.~Levenberg, D.~Mane,
  R.~Monga, S.~Moore, D.~Murray, C.~Olah, M.~Schuster, J.~Shlens, B.~Steiner,
  I.~Sutskever, K.~Talwar, P.~Tucker, V.~Vanhoucke, V.~Vasudevan, F.~Viegas,
  O.~Vinyals, P.~Warden, M.~Wattenberg, M.~Wicke, Y.~Yu, X.~Zheng, Tensorflow:
  Large-scale machine learning on heterogeneous distributed systems (2016).
\newblock \href {http://arxiv.org/abs/1603.04467} {\path{arXiv:1603.04467}}.

\bibitem{Akiba2019}
T.~Akiba, S.~Sano, T.~Yanase, T.~Ohta, M.~Koyama, Optuna: A next-generation
  hyperparameter optimization framework (2019).
\newblock \href {http://arxiv.org/abs/1907.10902} {\path{arXiv:1907.10902}}.

\bibitem{Diederik2014}
D.~P. Kingma, J.~Ba, Adam: A method for stochastic optimization (2014).
\newblock \href {http://arxiv.org/abs/1412.6980} {\path{arXiv:1412.6980}}.

\bibitem{Chen2019}
Y.-A. Chen, L.~P. Tripathi, T.~Fujiwara, T.~Kameyama, M.~N. Itoh, K.~Mizuguchi,
  \href{https://doi.org/10.3389/fgene.2019.00934}{The {TargetMine} data
  warehouse: Enhancement and updates}, Frontiers in Genetics 10 (10 2019).
\newblock \href {https://doi.org/10.3389/fgene.2019.00934}
  {\path{doi:10.3389/fgene.2019.00934}}.
\newline\urlprefix\url{https://doi.org/10.3389/fgene.2019.00934}

\bibitem{Kuna2019}
L.~Kuna, J.~Jakab, R.~Smolic, N.~Raguz-Lucic, A.~Vcev, M.~Smolic,
  \href{https://doi.org/10.3390/jcm8020179}{Peptic ulcer disease: A brief
  review of conventional therapy and herbal treatment options}, Journal of
  Clinical Medicine 8~(2) (2019) 179.
\newblock \href {https://doi.org/10.3390/jcm8020179}
  {\path{doi:10.3390/jcm8020179}}.
\newline\urlprefix\url{https://doi.org/10.3390/jcm8020179}

\bibitem{Lee2019}
C.~Y. Lee, Y.-P.~P. Chen,
  \href{https://doi.org/10.1016/j.drudis.2019.03.003}{Machine learning on
  adverse drug reactions for pharmacovigilance}, Drug Discovery Today 24~(7)
  (2019) 1332--1343.
\newblock \href {https://doi.org/10.1016/j.drudis.2019.03.003}
  {\path{doi:10.1016/j.drudis.2019.03.003}}.
\newline\urlprefix\url{https://doi.org/10.1016/j.drudis.2019.03.003}

\bibitem{Joseph2013}
P.~Joseph, C.~Umbright, R.~Sellamuthu,
  \href{https://doi.org/10.1002/jat.2861}{Blood transcriptomics: applications
  in toxicology}, Journal of Applied Toxicology (2 2013).
\newblock \href {https://doi.org/10.1002/jat.2861}
  {\path{doi:10.1002/jat.2861}}.
\newline\urlprefix\url{https://doi.org/10.1002/jat.2861}

\bibitem{drugmatrix}
D.~L. Svoboda, T.~Saddler, S.~S. Auerbach, An overview of national toxicology
  program’s toxicogenomic applications: Drugmatrix and toxfx, in: Advances in
  Computational Toxicology, Springer, 2019, pp. 141--157.

\end{thebibliography}
